# Chirality-Dependent Kinetics of Single-Walled Carbon Nanotubes from Machine-Learning Force Fields

Sida Sun, Shigeo Maruyama, Yan Li*

**ABSTRACT:** The origin of the chirality of single-walled carbon nanotubes (SWCNTs) has been a long-standing dispute. Molecular dynamics (MD) simulations driven by machine-learning force fields (MLFF), which can study the interface dynamics under near ab-initio accuracy, provides a powerful technique to reveal the formation mechanism of SWCNTs. Here, we develop a cobalt-carbon MLFF and perform growth simulations on a cobalt catalyst to investigate the chirality preference of the growth of SWCNTs under the vapor-liquid-solid (VLS) regime. Through microkinetic modeling, we reproduce the observed growth and defect kinetics, demonstrating their dependence on the chirality. It is observed that while the initial chirality assignment is likely related to the configurational degeneracy of the nanotube caps, pentagon defects immediately form and resolve after nucleation. Such processes, which we name as diameter control mechanisms, not only control the diameter toward an optimum but also shift the chirality distribution drastically. Our work therefore offers a microkinetic modeling workflow for the chirality-dependent kinetics of the SWCNTs, highlighting the important contribution of the defect kinetics to the chirality origination.

After three decades of the discovery of single-walled carbon nanotubes (SWCNTs)[1], the origin of their chirality still remains a mystery. Due to the lack of *in situ* experimental methods to characterize the nanotube-catalyst interface structure and its dynamic evolution during the growth and nucleation process at atomic scale, theoretical methods ought to take the main role in revealing the mechanism. The first theoretical works on the possible origin of the chirality appeared in 2006, in which Robertson *et al.* proposed that the epitaxy between the edge of certain nanotube caps and catalyst surfaces can generate chirality preference[2,3]. Since then, many theories have been proposed, mainly regarding the dependence of the nucleation probability and the growth rate on the chirality.

Early formulations on these two quantities are proposed by Yakobson *et al.* They considered the nucleation within classical nucleation theory[4–6], writing the energy of the critical nucleus as $G = G_{\mathrm{cap}} + \Gamma$ and the nucleation probability as $\exp(-G/k_{\mathrm{B}}T)s/Z$, where $s$ is a degeneracy factor related to possible number of caps, etc., and $Z$ is the partition function. The $G_{\mathrm{cap}}$ term is the elastic energy of the nanotube cap, which is independent on the nanotube diameter as they proposed[7]. Therefore, the chiral angle preference in the nucleation originates only from the interfacial energy $\Gamma$ of the edge-catalyst interface, which can be expressed as $\Gamma = N_{\mathrm{A}} \varepsilon_{\mathrm{A}} + N_{\mathrm{Z}} \varepsilon_{\mathrm{Z}} + \Gamma_{\mathrm{mix}}$, where A and Z represents armchair and zigzag edges, and $\Gamma_{\mathrm{mix}}$ is a correctio term for the mixing of these A- and Z-edges. Their initial model of the growth kinetics is the screw dislocation theory[4,8], linking the number of "kinks" at the edge to the growth rate. Their model favors $(n, n-1)$ chiralities on liquid catalysts and $(2m, m)$ chiralities on solid catalysts. They further switched to more sophisticated models based on density-functional theory calculations and Kinetic Monte Carlo simulations.

Later, Bichara *et al.* developed their thermodynamic models[9–11], where they added the edge configurational entropy of nanotubes onto the edge energy. From this model, they constructed phase diagrams that identify the stable chiralities under different parameters. Ding *et al.* proposed another model in which the nucleation can be kinetically controlled[12] — on liquid catalysts, the chirality determination is random; but on solid catalysts, the different surface sites can possibly induce chiral selectivity.

On the other hand, molecular dynamics (MD) simulations can function as a computational microscope to observe the atomic motion during the growth process. Maruyama *et al.* first realized the few-defects and defect-free growth simulations using classical force fields[13,14]. Very recently, they together with Hedman *et al.* introduced machine-learning force fields (MLFFs) of the iron-carbon system into the simulations, which present much improved accuracy[15,16]. They confirmed that the nanotube edge actually displays diverse patterns during growth, which is overlooked in early theories.

While the dynamics has been observed in simulations, it is more important to develop a theoretical model which can rationally predict the growth outcomes. In this work, with a home-developed MLFF, we simulate the growth of SWCNTs on cobalt catalysts because cobalt-based catalysts have shown great success in chirality-selective growth of SWCNTs[17–19]. We propose a new representation for faithful edge pattern statistics and performed microkinetic modeling of the growth and defect kinetics at the nanotube-catalyst interface using the statistics. Through these results, we concluded that the chirality distribution might depend largely on the defect kinetics at the nanotube-catalyst interface.

## RESULTS AND DISCUSSION

**SWCNT edge patterns and reaction networks.** We propose a representation based on Brinkmann *et al.*'s work[20] to enable the rigorous enumeration of all edge patterns. By definition, the edge patterns can be partitioned by the number of bonds connected to the edge atoms, or equivalently the degree of edge vertices. Denoting deg-2 and deg-3 vertices as "2" and "3", the edge patterns can be partitioned into the form of $n$ pairs of "23" and $m$ pairs of "32", and it is trivial that this partition is unique for chiral SWCNTs. To simplify the notation, we denote the "23" pairs with "0" and the "32" pairs with "1" in our following discussions. Such partitioned edge patterns thus form circular strings of "0"-s and "1"-s, which are called $(n+m)$-length binary necklaces in combinatorics. Taking $(n,m) = (6,3)$ as an example (Figure 1a), with six "0"-s (violet beads) and three "1"-s (blue beads) on a necklace, there are ten unique edge patterns according to Polya enumeration theorem:

$$N_{\text{edge}} = \frac{1}{n+m} \sum_{d \mid \gcd(n,m)} \phi(d) \frac{((n+m)/d)!}{(n/d)!\,(m/d)!}$$

where $\gcd(n,m)$ denotes the greatest common divisor of $n$ and $m$, and $\phi(d)$ is the Euler totient function. Generating an exhaustive list of these edge patterns can be done by simply permutating the necklace and canonicalizing them to their lexicographical minima. For example, if an observed edge pattern for (6,3) chirality reads "100000011", then by rotating the longest consecutive sequence of six "0"-s to the front, the canonicalized edge pattern "000000111" is recovered.

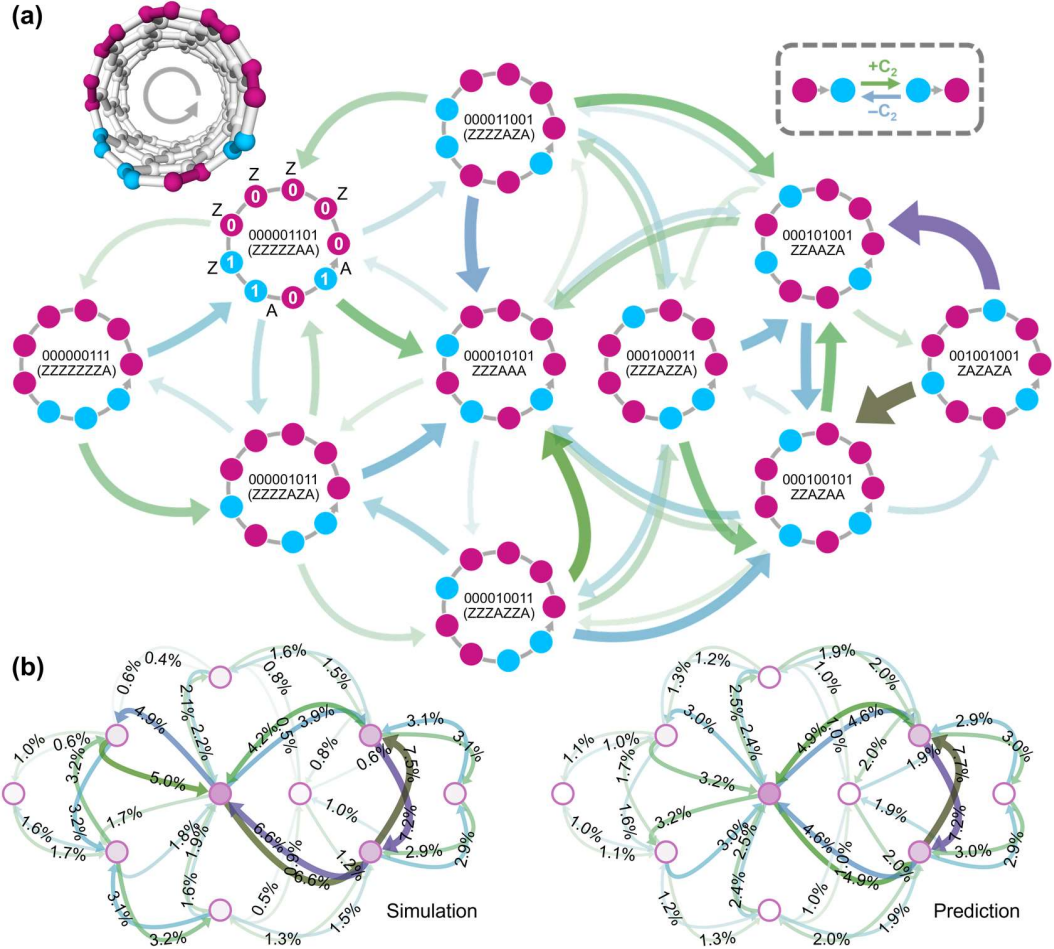

**Figure 1.** "0-1" and "A-Z" edge pattern representations and reaction networks of a (6,3) single-walled carbon nanotube (SWCNT). **(a)** Construction of the reaction network. The edge atoms and bonds are colored by the type of atomic pair they are in, the pair "23" ("32") in violet (blue). This notation is further simplified by denoting "23" ("32") with "0" ("1") or violet (blue) beads on a necklace. After assigning "10" to A-edges, all "0"-s and "1"-s left are Z-edges. However, if both "0"-s and "1"-s are left, such "A-Z" representation may not be unique and is displayed in parentheses. All possible reactions are $C_2$ addition and etching over anti-armchair sites or A-edges, which can be represented as "01" → "10" and "10" → "01", respectively. The green (blue) arrows show the possible $C_2$ addition (etching) reactions, and their arrow widths denote the rates which are fitted to the statistics from the machine-learning force field-driven molecular dynamics (MLFF-MD) simulation. **(b)** Reaction incidences and the steady-state distributions from the MLFF-MD simulation (left) and the model prediction (right).

Such "0-1" representation is required to produce faithful statistics since it fills the loophole of the traditional armchair- (A-) and zigzag- (Z-) edges representation, which is not an isomorphism from the "A-Z" strings to the atomic structures. For edge patterns with $i < m$ A-edges, the $i$ anti-armchair sites can be nonadjacent to their A-edges, and the Z-edges can align differently with respect to the tube axis. This is more obvious in the "0-1" representation, where an A-edge equals to the pattern "10", and after excluding the "10" patterns, all the "0"-s and "1"-s left in the pattern are Z-edges. If such Z-edges are not all "0"-s or all "1"-s, the edge pattern might not be well defined in the "A-Z" representation. Again using (6,3) as an example, in Figure 1a there are six edge patterns with less than $m = 3$ A-edges, their "A-Z" representation denoted in parentheses. Among them, "000011001" and "000001011" both corresponds to "ZZZZAZA", while "000010011" and "000100011" both corresponds to "ZZZAZZA". By adopting the "0-1" representation, these ambiguities can be avoided.

We then explore the interactions between these edge patterns. Etching a $C_2$ molecule from an A-edge can be expressed as "10" → "01", and the reverse process "01" → "10" is the addition of a $C_2$ molecule to an anti-armchair site. Since all possible reactions that do not change the edge pattern length $(n + m)$ are C$_2$ addition and etching, we can find all the A-edges and anti-armchair sites in the enumerated edge patterns and perform the reactions on it, thus constructing a pair of addition and etching reaction networks as two strongly connected directed graphs. Note that the edges with lengths longer than $(n + m)$ (referred to as anomalous edges hereafter), which are crucial for the growth of near-zigzag SWCNTs, are temporarily not considered. Therefore, chiralities such as $(n, 0)$ and $(n, 1)$ are effectively excluded from our discussions.

Having obtained the reaction networks, we should be able to predict the steady-state distribution $\mathbf{p}_\infty$ of the master equation $d\mathbf{p}_t/dt = R\mathbf{p}_t$ by $d\mathbf{p}_\infty/dt = 0 = R\mathbf{p}_\infty$. Here the transition matrix $R = Q_+ + Q_- - \text{diag}\{\mathbf{1}^T(Q_+ + Q_-)\}$ is built from the two networks, $Q_+$ for C$_2$ addition reactions and $Q_-$ for C$_2$ etching reactions, and their strongly connectedness guarantees the uniqueness of $\mathbf{p}_\infty$[21]. Nonetheless, our first attempt is to avoid the rate constants and approximate the $\mathbf{p}_\infty$ with Boltzmann distributions of the edge energies. Such edge energies are approximated as the sum of the atomic energies of all edge atoms, which are readily available from the MLFF. However, the total edge energies display fluctuations magnitudes larger than $k_B T$ for each pattern, and their averages do not follow the Boltzmann distribution (Figure S1). We thus suggest that the edge energies might not be defined in such a way. It should be more convenient to work directly on the environment-dependent reaction rates rather than formulate an energy functional, especially when the edge pattern breaks the $(n + m)$ length constraint.

Consequently, we continue to build the full transition matrix by assigning rate constants based on simple assumptions. We beforehand define the excess chemical potential $\Delta\mu = \mu_{\text{SWCNT}} - \mu_\text{C}$, where $\mu_{\text{SWCNT}}$ is the carbon chemical potential in an SWCNT, and $\mu_\text{C}$ is the carbon chemical potential in the catalyst. We first assume that, at zero $\Delta\mu$, the reactions that do not change the edge pattern have equal addition and etching rates $k_0 = A_0 \exp(-\beta \Delta G^\ddagger)$, where $\beta^{-1} = k_B T$, $A_0$ is the preexponential factor, and $\Delta G^\ddagger$ is the activation free energy in the transition state theory. At finite $\Delta\mu$, they can be multiplied by a factor of $\lambda_\mu^{\Delta i_\mu} = \exp(-\Delta i_\mu \beta \alpha(2\Delta\mu))$, where $\Delta i_\mu = 1$ denotes addition, $\Delta i_\mu = -1$ denotes etching, and $0 \leq \alpha \leq 1$ is the transition state coordinate stated in Bell-Evans-Polanyi principle. We further assume that, for reactions that change the edge pattern, their rate constants are multiplied by modifiers depending on the surrounding atoms of the reaction site. Currently, we consider up to four neighboring atoms on each side of the site, and the modifiers are determined by the change of the number of A-edges $i_A$ and A|Z junctions $i_{A|Z}$ in the local structure during the reaction. The modified rate of the reaction is thus $k_0 \lambda_\mu^{\Delta i_\mu} \lambda_A^{\Delta i_A} \lambda_{A|Z}^{\Delta i_{A|Z}}$ where $\lambda_A > 1$ and $\lambda_{A|Z} \lesssim 1$, thus being a four-parameter model. The intuition behind this assumption is that A-edges have better contact with the catalyst than Z-edges due to the absence of dangling bonds, and A|Z junctions make the edge "jagged" which result in poor contact. Further elaborations to this concept are presented in the Methods section of the Supporting Information.

We then put this growth model to a quick test by fitting it to a steady-state edge pattern distribution with (6,3) chirality, of which the MD simulation setup is described in the next section. Since its reaction networks in Figure 1a does not contain self-loops, it should be ideal for us to compare the predictions to the reaction incidence statistics (Figure 1b, other chiralities in the Appendix I of the Supporting Information). It can be seen that this analytical growth model, albeit simple, accounts for the underlying dynamics decently.

**Edge pattern statistics from MLFF-driven MD simulations.** Statistical analysis of the nucleation and the growth process requires hundreds of simulations which can extend to microsecond timescales, so it is crucial to trade off accuracy against efficiency. For our cobalt-carbon MLFF, we choose the DeepPot-SE model as implemented in the DeepMD-kit package[22-24] among available codes, noting its exceptional parallel performance for GPU-based MD simulations. To accelerate the development of this MLFF, we employ a home-built active learning workflow similar to existing methods like DP-GEN[25], which enable us to construct the MLFF in a data-efficient way. Starting from an initial training set with perturbed structures of different cobalt and carbon allotropes, we obtain the final dataset containing 6,523 structures with 529,211 atoms (Figure S2). To test the generalization ability of the model architecture, 11-fold cross-validation is performed on the final dataset, with mean energy and force RMSE of 8.4 meV / atom and 0.28 eV / Å, respectively. Physically motivated validations of the MLFF including equations of states of carbon and cobalt allotropes and melting point simulations of bulk elemental cobalt are also performed (Figure S3-S5).

Having obtained the MLFF, we then carry out large-scale MD simulations of the deposition process. We use the icosahedral Co$_{55}$ catalyst throughout the work and deposit atomic carbon on its surface with a slow rate of one carbon atom per 2 ns, after a faster deposition with a rate of one atom per 0.1 ns to enhance the nucleation. Trajectories are collected at a 20 ps interval to save disk space. Nucleated nanotube caps are allowed to freely evolve until they form five or more defect-free hexagon layers, or a defect trapped in the nanotube wall is at least five hexagons away from a cap-forming pentagon or heptagon. For nanotube caps with diameters under 1.0 nm, cap-forming pentagons are usually less than two hexagons away from each other. Therefore, the curvature at the stopping location should be close to that of its corresponding bulk SWCNT.

We perform all simulations at 1500 K, and a total of 210 cases are collected according to the above criterion. Among these cases, 34 cases with various chiralities are allowed to evolve for at least ten layers of defect-free hexagons (referred to as "ten-layer" trajectories hereafter) while being defect-free, their diameters spanning from 0.61 nm to 0.95 nm (Figure 2a and Figure 3a). Excluding near-zigzag SWCNTs of which $m = 0, 1$ and SWCNTs with diameters larger than 0.9 nm, we perform edge pattern statistics on the 20 cases left and fit our growth model to these statistics, the (6,3) trajectory in the previous section included. The constant deposition rate used in our simulations can actually relate to a constant excess chemical potential $\Delta\mu < 0$, since they both should correspond to a constant growth rate. Still, the $\Delta\mu$ is not the same for different chiralities, and the base rate constant $k_0 = A_0 \exp(-\beta \Delta G^\ddagger)$ is likely not the same either, and we expect them to depend on the diameter of the SWCNT. Although these two quantities barely affect the edge pattern distribution, we can extract them by fitting the model with another two observables: the total reaction incidence $f = \mathbf{1}^T(Q_+ + Q_-)\mathbf{p}_\infty$ and the growth rate $r = \mathbf{1}^T(Q_+ - Q_-)\mathbf{p}_\infty$. Then the total objective function $J[k_0, \lambda_\mu, \lambda_A, \lambda_{A|Z}]$ should be

$$J = -\sum_j p_{\infty,j} \ln \hat{p}_{\infty,j} + w_1(f - \hat{f})^2 + w_2(r - \hat{r})^2$$

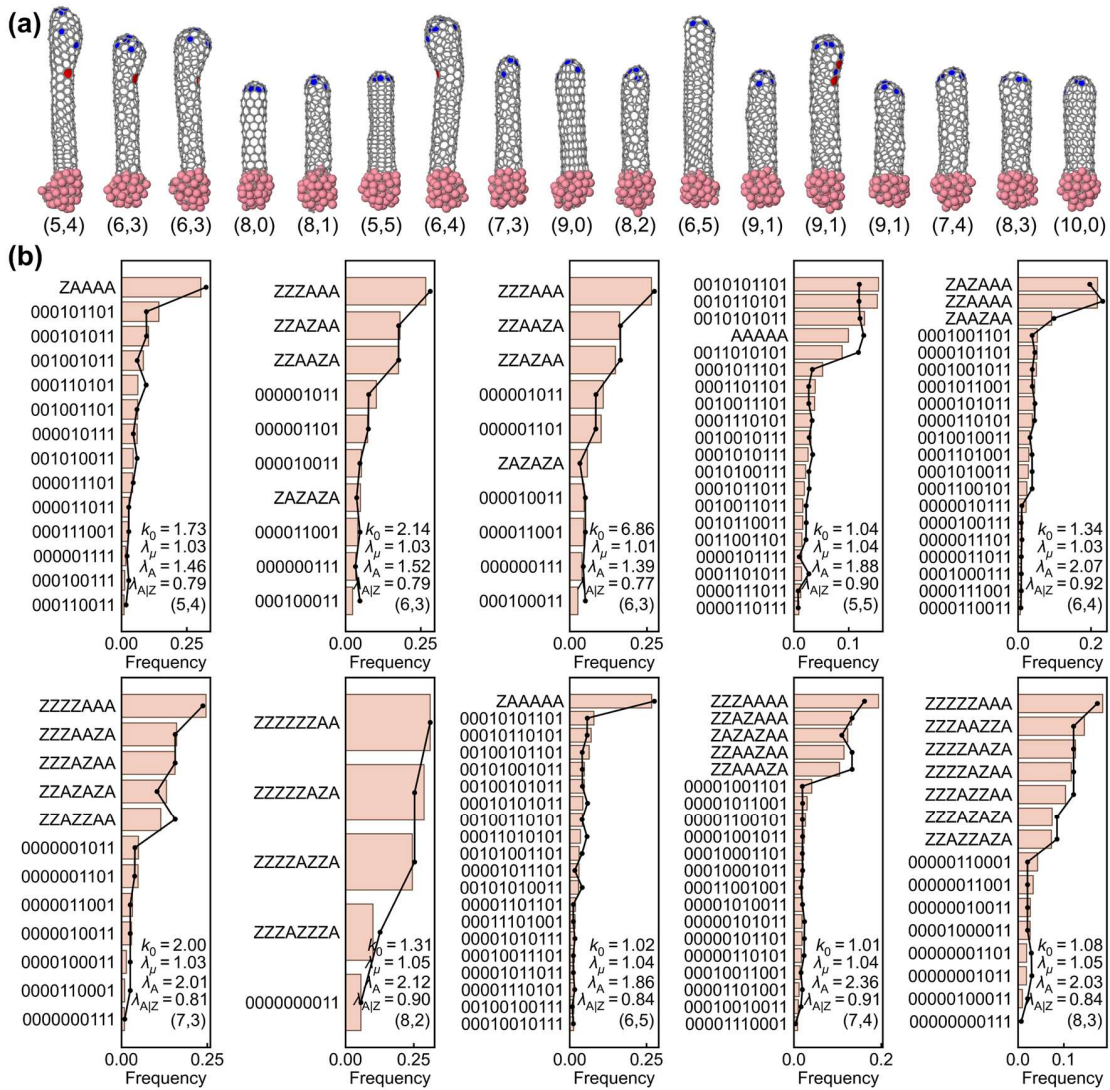

**Figure 2.** Edge pattern statistics and predictions from the growth model. (**a**) The 17 selected single-walled carbon nanotubes (SWCNTs) with smaller diameters among the 34 selected ones grown on the $Co_{55}$ catalyst using our developed machine learning force field (MLFF). These SWCNTs are allowed to evolve for at least ten layers of defect-free hexagons for detailed analysis. Pink (grey) spheres are cobalt (carbon) atoms, and pentagons (heptagons) are colored blue (red). (**b**) Edge pattern statistics (bars) derived from 10 of the above trajectories along with the fitted predictions from the growth model (lines with markers). The handedness is unified for clarity. Edge patterns are represented in the "A-Z" representation if possible.

where hatted variables denote predictions from the model, $r = 0.25$ $ns^{-1}$ is the target growth rate or correspondingly the deposition rate counted in $C_2$ dimers, $f$ is the target total reaction incidence estimated by counting changes in edge patterns, and $w_{1,2}$ are weight factors set to 1 ns. The estimated total reaction incidences are rescaled by letting the net addition frequencies match the growth rate of 0.25 $ns^{-1}$ so that the insufficient sampling is alleviated.

We display the most occurring edge patterns in Figure 2b and Figure 3b along with the fitted predictions from the growth model. As can be seen, the model describes the distributions of small diameter SWCNTs with $d_t$ less than ~ 0.82 nm very well, and $(n, 2)$ SWCNTs included in the statistics are also well fitted regardless of their near-zigzag nature and large fraction of anomalous edges. It can be seen that edge patterns with rotational symmetries, such as "ZAZAZA" for (6,3) and "ZAAZAA" for (6,4), occur much less often than their asymmetrical counterparts with the same number of Z- and A-edges. This phenomenon, described as the "conventional interface" tends to be "kinetically unstable" by Penev et al.[5], turns out to be of entropic origin, as these edges have less rotational degeneracy than their counterparts ($(n + m)/d$ versus $(n + m)$, where $d$ is the highest order of the rotational symmetry of the edge pattern). This is also evident from the rate constants of the reaction networks, as rate constants pointing outwards from these edge patterns are all multiplied by the factor $d$, regardless of adding or etching.

Still, it can be seen that the quality of fit decreases with the diameter. We thus suspect that the simulation time is not enough or the sampling interval is too long, since the number of edge patterns grows combinatorially with the chiral indices. For the two (7,6) SWCNTs with diameters of 0.88 nm, inconsistencies in the orders of the observed edge patterns between these two cases can be noticed, though the prevailing edge pattern "ZAAAAAA" remains the same. This is inevitable under limited computational resources, and we speculate that even a several-microsecond trajectory might not suffice when we need to distinguish between very similar probabilities of occurrence. Considering the simplicity of the growth model, further improving the statistics might be worthless unless a much more detailed model is developed.

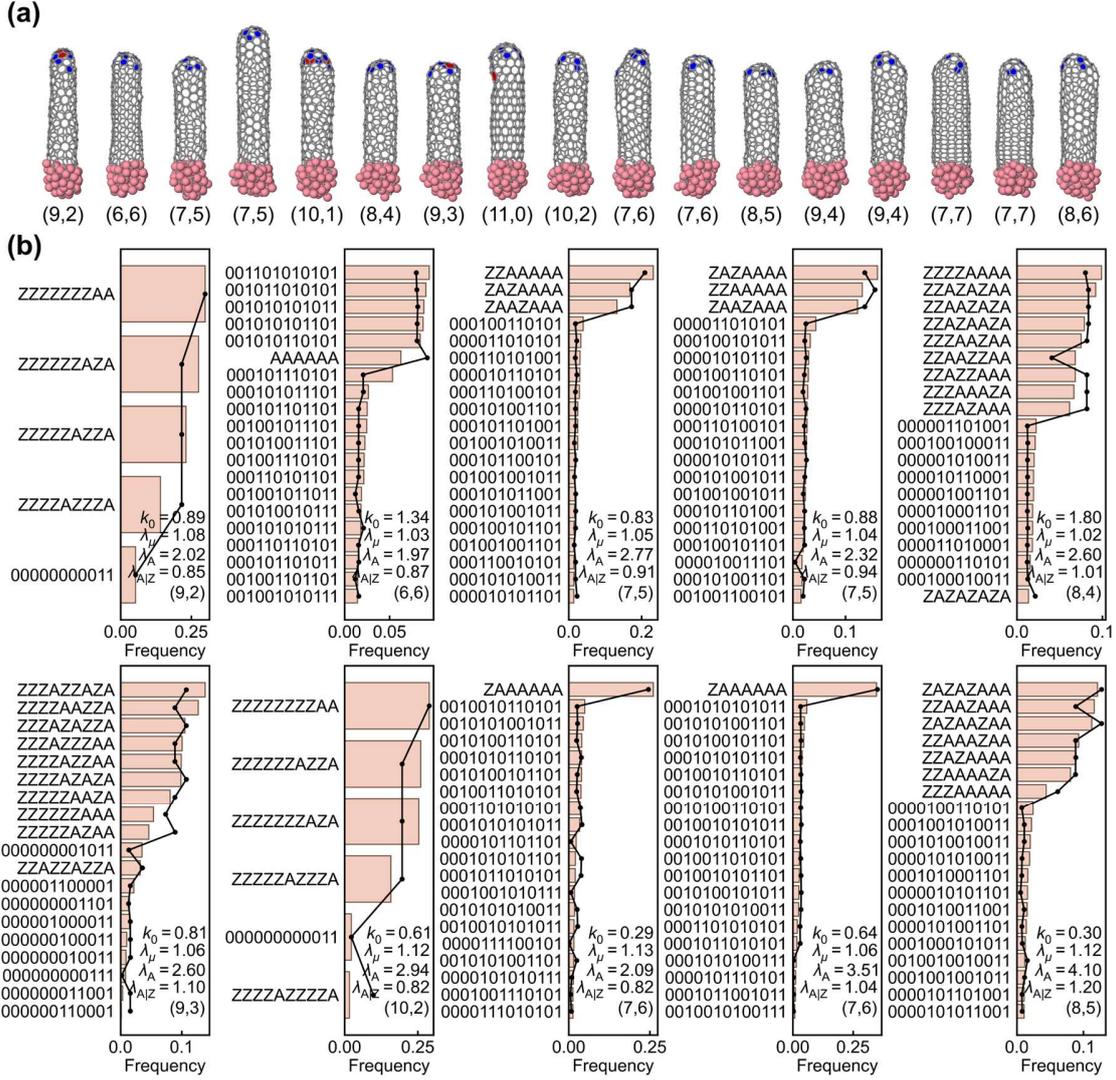

**Figure 3.** Similar to Figure 2, but for the structure and edge pattern distributions of the set of SWCNTs with larger diameters.

The fitted model parameters are shown in Figure 4, assuming $\alpha = 0.5$ in the Bell-Evans-Polanyi formalism: Since the reactant and the product have the same structure in our baseline reaction with the rate constant $k_0$, under $\Delta\mu = 0$ or equivalently $\Delta G = 0$ it should be natural to identify the transition state at the center of the reaction coordinate. Thus, we can define the excess chemical potential by $\Delta\mu = -k_B T \ln \lambda_\mu$, the A-edge free energy by $\Delta G_A = -2k_B T \ln \lambda_A$, and the A|Z junction free energy by $\Delta G_{A|Z} = -2k_B T \ln \lambda_{A|Z}$. As mentioned in the previous section, one would intuitively expect that $\Delta G_A < 0$ and $\Delta G_{A|Z} > 0$. As the diameter increases, however, it can be seen in Figure 4a that $\Delta G_{A|Z}$ approaches zero, and for some chiralities it even crosses over to negative values. Correspondingly, the edge patterns where A-edges and Z-edges are segregated become less and less prevailed over the other ones (Figure 2b and Figure 3b). This phenomenon can be attributed to insufficient statistics, but an alternative explanation is also possible: the growth-mode crossover from perpendicular to tangential[26–28]. For large diameter SWCNTs, their more tangential interface between the edge and the catalyst ensures better contact irrespective of the edge pattern, thus the A|Z junctions are not too penalized in terms of local free energy.

Another interesting observation is that for large diameter SWCNTs, their growth rate per anti-armchair site is inherently slower. Though the base rate constant $k_0$ is still likely underestimated, it is clear that $k_0$ decreases with the diameter, which can be seen from $\Delta G^\ddagger = -k_B T \ln k_0 + k_B T \ln A_0$ in Figure 4b. This is more pronounced when represented in terms of hexagon layers per time unit, since the growth rate in $C_2$ units should be further divided with $(n+m)$. And it is natural since larger diameter naturally translates to smaller curvature, lower reactivity, and thus smaller $k_0$ or higher $\Delta G^\ddagger$. Therefore, achieving the same growth rate as the small diameter SWCNTs, the excess chemical potential ($\Delta\mu < 0$) applied to large diameter SWCNTs is lower. This may affect the observed chirality distribution, which we will discuss in the next section.

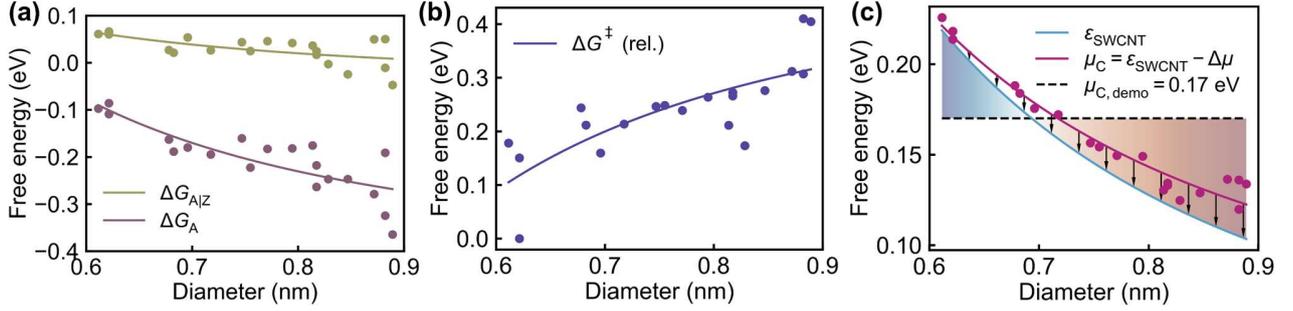

**Figure 4.** Thermodynamic quantities derived from the fitted models. (**a**) A-edge free energies and A|Z junction free energies. (**b**) Relative activation free energy of the baseline reaction. (**c**) Chemical potential (violet line) and excess chemical potential (arrows), showing that all diameters are preferred in the simulations. $\varepsilon_{SWCNT}$ (blue line) is the energy of the atoms in single-walled carbon nanotubes (SWCNTs) which is used as an approximation to the chemical potential. The dashed line and the shades demonstrate the harsher excess chemical potential experienced by SWCNTs with different diameters under a constant chemical potential. All solid lines in (a-c) are fitted against $y = a_0 + a_1/d_t^2$.

**Chirality preference from defect kinetics.** Regarding the 210 MD trajectories collected at 1500 K mentioned above, 209 of them developed towards 42 different chiralities at five hexagon layers as in Figure 5a, displaying a very broad distribution. We speculate that this is because all chiralities are intrinsically preferred, i.e., $\Delta\mu < 0$ in terms of excess chemical potential, which may not be the case when the external chemical potential $\mu_C$ is constant. At constant $\mu_C$, the excess chemical potential is approximately $\Delta\mu = c_0/d_t^2 - \mu_C$ when measured from the energy of a carbon atom in graphene, where $d_t$ is the diameter of the SWCNT, and $c_0$ is the out-of-plane elastic constant of graphene[29]. In Figure 4c, we use a $\mu_{C,demo} = 0.17$ eV to demonstrate how the experienced $\Delta\mu$ in experiments differs from that in our simulations. For nanotube caps with too small diameters, even if they successfully nucleate from the excessively supersaturated carbon dissolved in the catalyst, they actually experience a $\Delta\mu > 0$ at equilibrium. Therefore, they should be etched, re-nucleate to larger nanotube caps, or form heptagon defects to increase their diameters. On the other hand, nanotube caps with too large diameters receive $\Delta\mu$ much lower than that in simulations. They may grow too fast to heal the extra pentagons formed at the interface, encapsulating them and turning into conical structures with shrinking diameters. We therefore expect a much narrower diameter distribution in such conditions.

Nonetheless, the two scenarios just mentioned are actually observed in our simulations since our $\mu_C$ is still kind of "adaptive". There are two of our SWCNTs with ultra-small diameters displaying the first scenario, their chiralities being (6,2) and (7,2). After forming multiple hexagon layers, their interfaces suddenly become tilted with one side of the wall being etched, increasing the circumference of the interface. For the second scenario there are even more examples. We can collect the last defined chirality after the chirality has been once defined for all 210 nanotube caps, as displayed in Figure 5b. Comparing this "zero-layer" distribution to the "five-layer" one, it can be seen that the "zero-layer" distribution is apparently consistent with the kinetic nucleation hypothesis by Xu et al., where the nucleation probability hardly depends on the chiral angle[12]. Actually, this "zero-layer" distribution may serve as the boundary of the nucleation phase in the strict sense, as the chirality distribution at the preferred diameter roughly agrees with the number of the nanotube caps containing only pentagons and hexagons (Figure S6). However, the "five-layer" distribution display a huge shift mainly toward smaller diameters, and this is mostly due to the subsequent formation of shrinking cones after the strict nucleation phase. We thus suggest that the nanotube-catalyst interface should not be that "smart" to know how many pentagons are encapsulated in the SWCNT — The chemistry should be local at the interface, depending only on the edge pattern.

As for the shrinking cones, obviously they cannot shrink forever. There are two basic mechanisms to increase the interface circumference, which we name as diameter control mechanisms. The type-I mechanism is to try healing the defect by etching on the side of the wall and forming a tilted interface, just like the aforementioned ultra-small diameter SWCNTs. This may not succeed though, as the etching can fail to target the defect, leaving it in the wall permanently. The type-II mechanism is to form a heptagon defect to compensate for the large curvature, switching the chirality in progress. Demonstrative snapshots for these mechanisms are displayed in Figure 5c-e. Also, the formation of the cap-forming pentagons (or the chirality-switching pentagons) can be regarded as controlling the diameter toward the smaller side, especially when there are insufficient pentagons, and thus the cap is expanding in diameter but cannot expand forever.

Apart from providing such insights into diameter control mechanisms, the MD simulations enlighten us to realize the connection between the chirality preference and the defect kinetics, as diameter control is also a form of chirality control. We thus proceed to analyze the kinetics of the pentagon defects of all the 34 "ten-layer" trajectories. As stated above, all the "ten-layer" trajectories selected are growing steadily without defects in their hexagon walls throughout their timespans. However, in other cases, defects can stay until they are deep inside the nanotube wall, and finally lead to diameter control mechanisms and possibly chirality switching. Previously, Hedman et al. fitted their observed distribution of defects with a power-law function, i.e., $\varphi_\tau = g_0 \tau^{-g_1}$, where $g_0$ and $g_1$ are constants[15]. By fitting it against the mean distribution from the "ten-layer" trajectories, we do reproduce this behavior, obtaining $g_1 \sim 2$ (Figure 6a). This makes us question the actual mechanism of defect healing, since no simple rate law can produce this behavior. Basically, most of the nascent pentagon defects can be classified into three types by the number of its surrounding hexagons $l = 2,3,4$ (Figure 6b). From our statistics in Figure 7a, these three types of defects decay differently. $l = 2$ defects are formed over consecutive Z-edges, can be healed quickly, and can seldom lead to further reactions; $l = 3$ defects are hosted on anti-armchair sites, usually being intermediates of $C_2$ addition reactions; $l = 4$ defects are hosted on anti-zigzag sites in anomalous edges. In our system, $l = 4$ defects heal much slower than $l = 3$ defects, resulting in a larger chance of encapsulation, which is probably due to the relatively perpendicular contact lifting the defect out of the nanotube-catalyst interface. Therefore, the observed power-law decay is actually composed of multiple processes with different nature.

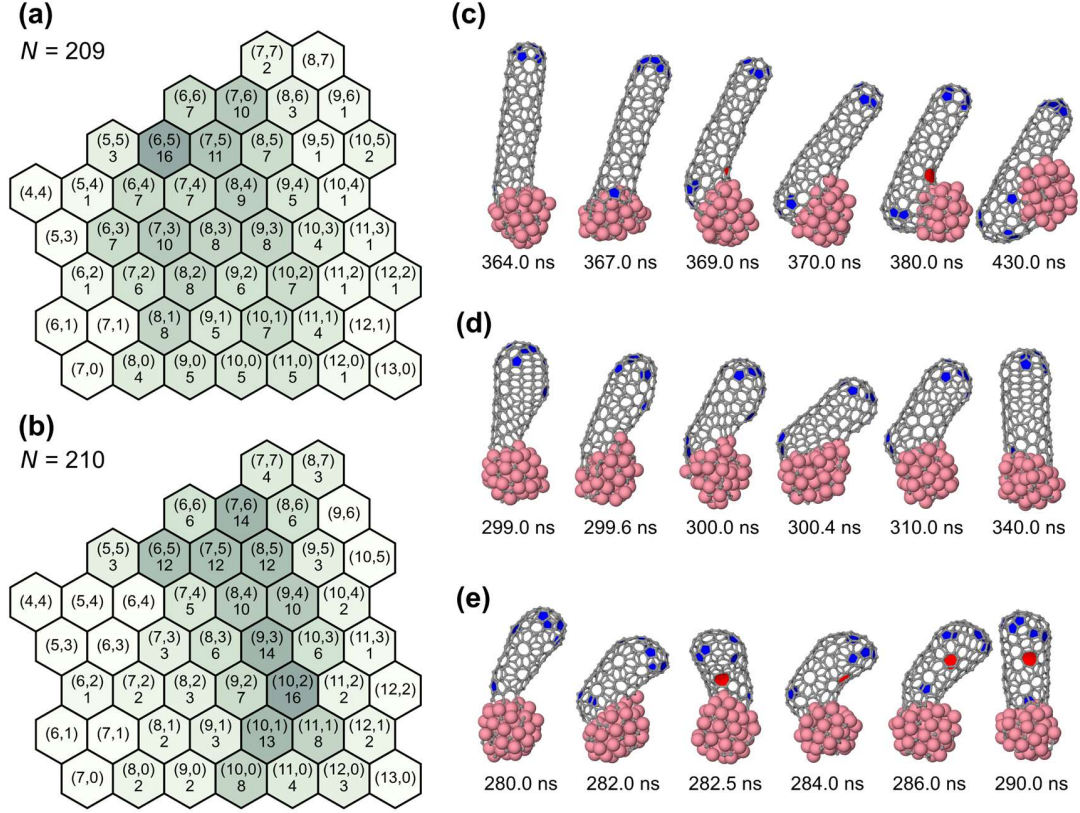

**Figure 5.** Chirality distributions and observed diameter control mechanisms that resolve small circumferences of the nanotube-catalyst interface. (**a**) "Five-layer" distribution, taken at five defect-free hexagon layers for all cases, except one case that has finally detached from the catalyst. Three chiralities are not included and are listed as follows: (12,4): 1, (13,1): 1, (13,4): 2. (**b**) "Zero-layer" distribution, taken at the end of the "strict" nucleation phase. (**c**) Type-I mechanism with (7,2) chirality. (**d**) Type-I mechanism with (7,5) chirality. The extra pentagon in the wall is successfully healed. (**e**) Type-II mechanism of a transition from (likely) (8,4) to (8,3). The encapsulated pentagon is not healed, and a heptagon is formed to compensate for it. Pink (grey) spheres are cobalt (carbon) atoms, and pentagons (heptagons) are colored blue (red) in (c-e).

Taking these observations into account, a crude, linear model of pentagon defects without considering the geometries should be:

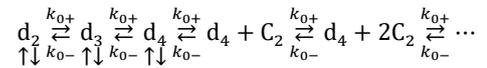

where $d_l$ denotes the defect surrounded by $l = 2,3,4$ hexagons, upward (downward) arrows denote defect generation (healing), and $k_{0+} \sim k_{0-} \sim k_0$ is the rate constant of hexagon addition or etching near equilibrium. The average defect healing rate constants of $l = 2,3,4$ defects for all chiralities are $k_{d2}$, $k_{d3}$, and $k_{d4}$, respectively, and the defect generation rate constants are temporarily ignored here. By solving this defect model numerically with rejection-free kinetic Monte Carlo simulations, the power-law behavior can be reproduced as in Figure 7b. For each initial state, we perform $10^6$ simulations with different random seeds, truncating the reactions to 20 $C_2$ additions and setting the lifetime limit to 500 ns so that infinite lifetimes are avoided. The lifetimes obtained are discretized to multiples of 20 ps in the histogram to match the sampling interval in MD simulations. The rate constants are fitted with Bayesian optimization, being $k_0 = 1.63$ ns$^{-1}$, $k_{d2} = 104$ ns$^{-1}$, $k_{d3} = 17.7$ ns$^{-1}$, and $k_{d4} = 3.63$ ns$^{-1}$. The value of $k_0$ is comparable to that from the growth model. The exact numerical distributions are also calculated alongside the histograms by integrating the master equation $d\mathbf{p}_{\tau,d}/d\tau = R_d \mathbf{p}_{\tau,d}$ and taking the time derivative as $\varphi_\tau = d(1 - \mathbf{1}^T \mathbf{p}_{\tau,d})/d\tau$, where the subscript d denotes quantities for this defect model. It can be seen that while at $\tau \to 0$ the observed power-law divergence of $\varphi_\tau$ is actually due to aliasing from finite-time sampling, the large $\tau$ behavior is consistent with the power law. We thus confirm that the chirality-switching pentagon defects arise from just a one-step side reaction between the normal $C_2$ addition and etching reactions.

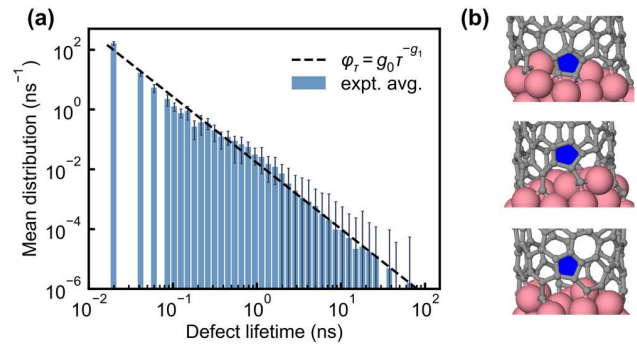

**Figure 6.** (**a**) Mean defect lifetime distribution of the 34 trajectories. Error bars denote the range, and the dashed line is fitted against a power-law function. (**b**) Three classes of pentagon defects classified by their number of shared edges $l = 2, 3, 4$ with the hexagons.

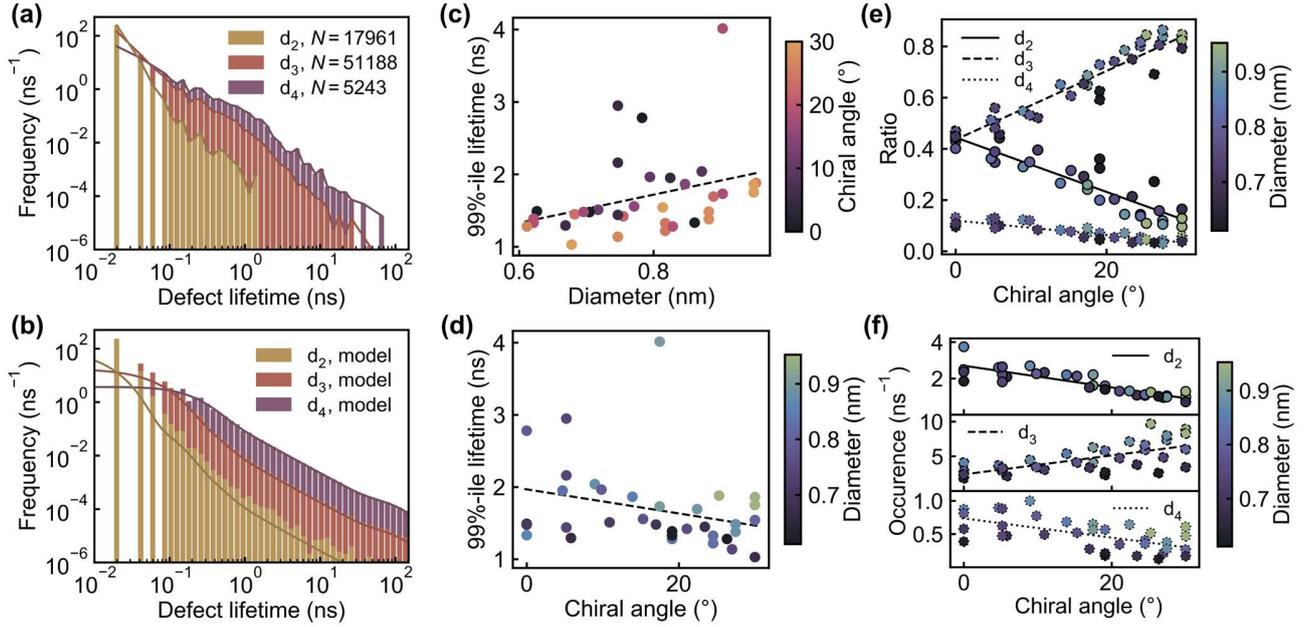

**Figure 7.** Statistics of the pentagon defects, including lifetimes and occurrences. (**a**) Lifetime distribution of the pentagon defects, classified by their surrounding hexagons $l = 2,3,4$ when they form. (**b**) Fitted predictions of the lifetime distributions from our defect model. Bars are results from the kinetic Monte Carlo simulations, and solid lines are the exact numerical results. (**c-d**) 99% percentile lifetime of the defects versus the diameters (c) and the chiral angles (d). Dashed lines are only intended as visual aids. (**e-f**) Ratio of the occurrences (e) and the raw occurrences (f) of different classes of defects versus the chiral angles fitted against $y = a_0 \theta + a_1$.

Finally, we analyze the chirality-dependent behavior of the defects to unveil their potential effects on the chirality distribution. It is obvious from the model that at a fixed temperature, larger diameter leads to lower $\Delta\mu$ or equivalently larger $\lambda_\mu$, consequently increasing the probability of chirality switching. Together with the inherently slower $k_0$ for large diameter SWCNTs, this is possibly reflected in the positive correlation between the defect lifetimes and the diameters (Figure 7c) and should be more pronounced at constant $\mu_C$. To explain the chiral angle distribution against near-zigzag SWCNTs at the optimal diameter, we consider the chiral angle dependence of the occurrence of $l = 3$ and $l = 4$ defects, since $l = 2$ defects hardly evolve into encapsulated defects as discussed above. For near-zigzag chiralities, their growth depends more on anomalous edges with anti-zigzag sites, hosting more $l = 4$ defects. As the chiral angle increases, more anti-armchair sites are available, thus $l = 3$ defects are favored over $l = 4$ defects (Figure 7e). Though the total occurrences of $l = 3$ and $l = 4$ defects in-crease with the chiral angle (Figure 7f), the faster healing rate of $l = 3$ defects in our system causes the negative correlation between the defect lifetimes and the chiral angles (Figure 7d), resulting in the preference toward near-armchair chiralities.

## CONCLUSION

Through our edge pattern formalism and our growth model, we successfully reproduce the VLS growth kinetics in our MLFF-driven MD simulations on cobalt catalysts, providing concrete evidence that $C_2$ addition and etching are the major contributing reactions except for near-zigzag SWCNTs. The thermodynamic parameters that govern the growth process are extracted, laying solid foundations for the operando modeling of the SWCNT formation.

More importantly, our simulations on cobalt catalysts reveal the plausible mechanisms of diameter evolution and control. The formation and resolution of defects drive the chirality distribution toward a steady state. In our growth model, the defect pentagons share similar origin with the cap-forming pentagons, and their kinetics is also dependent on the chirality. This prompts us to rethink the definition of nucleation, since defects may occur immediately after nucleating a conventional cap. While an underlying thermodynamic theory of such kinetics is not clear at the moment, they can definitely generate chirality preference. By applying our workflow to other elemental or alloy systems, one may discover different defect kinetics and potentially those leading to special chirality distributions. Future extensions to this workflow might involve the probability of chirality switching and growth termination per $C_2$ addition, therefore providing the decay exponents for Schulz-Flory-like macroscopic modeling[30,31]. Our present work thus paves the way for future studies on the growth kinetics and the rational design of catalysts for chirality-selective growth of SWCNTs.

## AUTHOR INFORMATION

### Corresponding Author

**Yan Li** – Beijing National Laboratory for Molecular Sciences, Key Laboratory for the Physics and Chemistry of Nanodevices, State Key Laboratory of Rare Earth Materials Chemistry and Applications, College of Chemistry and Molecular Engineering, Peking University, Beijing 100871, China
Email: yanli@pku.edu.cn

### Authors

**Sida Sun** – Beijing National Laboratory for Molecular Sciences, Key Laboratory for the Physics and Chemistry of Nanodevices, State Key Laboratory of Rare Earth Materials Chemistry and Applications, College of Chemistry and Molecular Engineering, Peking University, Beijing 100871, China

**Shigeo Maruyama** – Department of Mechanical Engineering, The University of Tokyo, Tokyo 113-8656, Japan

### Notes

The authors declare no competing financial interest.

## ACKNOWLEDGMENT

This work is supported by High-Performance Computing Platform of Peking University.